\documentclass[%
 reprint,
showpacs,preprintnumbers,
nofootinbib,
 amsmath,amssymb,
 aps,
 showkeys,
 showpacs
]{revtex4-1}

\usepackage{graphics}
\usepackage{bbold}
\usepackage{color}

\usepackage{array}
\usepackage{booktabs}
\usepackage{amsmath}
\usepackage{graphicx}
\usepackage{subfigure}

\usepackage{amsmath}
\usepackage{slashed}

\def\be		{\begin{eqnarray}}
\def\en		{\end{eqnarray}}
\def\nen	{\nonumber\end{eqnarray}}
\def\no		{\nonumber}
\def\lmo		{\left|}
\def\rmo		{\right|}
\def\lt		{\left(}
\def\rt		{\right)}

\def\hh		{\hspace{5 mm}}

\def\jp		{\ensuremath{J\!/\psi}}

\def\ov		{\ensuremath{\overline}}
\def\ee		{\ensuremath{e^+e^-}}

\def\bb		{\ensuremath{\mathcal{B}\overline{\mathcal{B}}}}
\def\b		{\ensuremath{\mathcal{B}}}
\def\aggg	{\ensuremath{\mathcal{A}^{ggg}}}
\def\agg	{\ensuremath{\mathcal{A}^{gg\gamma}}}
\def\ag		{\ensuremath{\mathcal{A}^{\gamma}}}

\def \br {{\rm{BR}}}

\begin{document}
\title{Strong and electromagnetic amplitudes of the $J/\psi$ decays into baryons and their relative phase}
\author{
\begin{normalsize}
Rinaldo~Baldini~Ferroli$^{1}$, Alessio~Mangoni$^{2,3}$, Simone~Pacetti$^{2,3}$, Kai~Zhu$^{4}$
\vspace{1 mm}\\
\vspace{0.2cm} 
\begin{normalsize} {\it
$^{1}$INFN Laboratori Nazionali di Frascati, I-00044, Frascati, Italy\\
$^{2}$Universit\`a di Perugia, I-06100, Perugia, Italy\\
$^{3}$INFN Sezione di Perugia, I-06100, Perugia, Italy\\
$^{4}$Institute of High Energy Physics, I-100049, Beijing People's Republic of China\\
}\end{normalsize}
\vspace{1.0cm}
\end{normalsize}
}

\begin{abstract}
The Feynman amplitude for the decay of the $J/\psi$ meson into baryon-antibaryon can be written as a sum of three sub-amplitudes: a purely strong, a purely electromagnetic and a mixed strong-electromagnetic. Assuming that the strong and mixed strong-electromagnetic sub-amplitudes have the same phase, the branching ratio of the decay contains an interference term that depends on the relative phase $\varphi$ between strong and electromagnetic sub-amplitudes. In this work we calculate this phase, by using an effective strong Lagrangian density and considering, as final states, pairs of baryons, $\mathcal{B}\overline{\mathcal{B}}$, belonging to the spin-1/2 SU(3) octet. Moreover, we obtain the purely strong, purely electromagnetic and mixed strong-electromagnetic contributions to the total branching ratio and hence the moduli of the corresponding sub-amplitudes, for each pair of baryons. Of particular interest is the mixed strong-electromagnetic contribution that, not only is determined for the first time, but it is proven to be crucial, in the framework of our model, for the correct description of the decay mechanism. Finally we use the purely electromagnetic branching ratio to calculate the Born non-resonant cross section of the annihilation processes $e^+ e^- \to \mathcal{B}\overline{\mathcal{B}}$ at the $J/\psi$ mass. By taking advantage from all available data, we obtain the relative phase between strong and electromagnetic sub-amplitudes: $\varphi = (73\pm 8)^\circ$.
\end{abstract}

\maketitle
\section{Introduction}
The decays of the $J/\psi$ meson into a baryon-antibaryon, \bb, final states proceed via strong and electromagnetic (EM) interactions. The Feynman amplitude can be written as a sum of three sub-amplitudes~\cite{3amplitudes}
$$
\mathcal{A}_{\bb} = \mathcal A^{ggg}_{\bb} + \mathcal A^{\gamma}_{\bb} + \mathcal A^{gg\gamma}_{\bb}  \,,
$$
where $\mathcal A^{ggg}_{\bb}$ is the purely strong, $\mathcal A^{\gamma}_{\bb}$ is the purely EM and $\mathcal A^{gg\gamma}_{\bb}$ is the mixed strong-EM sub-amplitude. At leading order, these sub-amplitudes are characterized by: a three-gluon ($ggg$), a one-photon ($\gamma$), and two-gluon-plus-one-photon ($gg \gamma$) propagators, respectively.
\\
Usually the mixed strong-EM contribution is not considered since it assumed to be negligible with respect to the strong and EM ones~\cite{gggamma-piccolo}. The calculation in the framework of QCD of this mixed contribution is a hard task because the hadronization process of the $gg\gamma$ into the final baryon-antibaryon pair does occur at a non-perturbative regime. 
\\
Furthermore it has been shown that in particular cases where the purely strong contribution is suppressed, e.g., in $G$-parity violating decays, the contribution of the mixed strong-EM term (related to $|\mathcal A^{gg \gamma}|$) cannot be neglected; see Ref.~\cite{Ferroli:2016jri} for the case of $J/\psi \to \pi^+ \pi^-$. 
\\
The branching ratio (BR) of the decays $J/\psi\to\bb$ contain the interference term between the sum of the purely strong and mixed sub-amplitudes, $\big(\aggg_{\bb}+\agg_{\bb}\big)$, for which we assume the same phase~\cite{Korner:1986vi}, and the purely EM sub-amplitude, $\ag_{\bb}$. In principle, the relative phase, called here $\varphi$, between these two contributions $\big(\aggg_{\bb}+\agg_{\bb}\big)$ and $\ag_{\bb}$ can be obtained by studying the $J/\psi$-resonance line-shape in processes as $e^+e^-\to J/\psi\to \bb$. 
\\
From the theoretical point of view, all sub-amplitudes should become real at a sufficiently high energy~\cite{Brodsky:1981kj,Chernyak:1984bm,Baldini:1996hc}, i.e., at a fully perturbative regime. On the other hand, it is quite difficult to establish whether such a regime, which entails maximum (positive or negative) interference, is attained already at the $J/\psi$ mass.
\\
A recent measurement~\cite{Ablikim:2012eu} of the relative phase $\varphi$, performed by exploiting the decay of the $J/\psi$ meson into nucleon-antinucleon, gave
$$
\varphi=(88.7\pm 8.1)^\circ\,,
$$
which is in agreement with the no-interference case.
\\
In this work we calculate the relative phase $\varphi$, the moduli of the sub-amplitudes $\mathcal A^{ggg}_{\bb}$, $\mathcal A^{gg\gamma}_{\bb}$ and $\mathcal A^{\gamma}_{\bb}$ and then their contributions to the total BR, using an effective strong Lagrangian density for the $J/\psi \to \bb$ decays, where the baryon-antibaryon pair belongs to the spin-$1/2$ SU(3) octet. Moreover, once the modulus of the EM sub-amplitude $\mathcal A^{\gamma}_{\bb}$ is known, $e^+ e^- \to\bb$ cross sections, at the $\jp$ mass, can be easily computed.
\section{Effective Lagrangian}
Let us consider the spin-1/2 SU(3) baryon octet, that in matrix notation reads
$$
\b=\begin{pmatrix}
\Lambda/\sqrt{6}+\Sigma^0/\sqrt{2} & \Sigma^+ & p\\
\Sigma^- & \Lambda/\sqrt{6}-\Sigma^0/\sqrt{2} & n\\
\Xi^- & \Xi^0 & -2 \Lambda/\sqrt{6}\end{pmatrix} \,.
$$
Since the $J/\psi$ meson is an SU(3) singlet state, the leading order Lagrangian density for the decay $J/\psi \to\bb$ should have the invariant form~\cite{Zhu:2015bha}
$$
\mathcal L^0 \propto {\rm Tr}\left(\bb\right) \,.
$$
Terms describing SU(3) symmetry breaking effects can be included to obtain a more complete Lagrangian density. We consider, in particular, two types of symmetry breaking sources: the quark mass difference and the EM interaction. The first one can be parametrized by introducing the ``spurion" matrix~\cite{Zhu:2015bha,Haber:1985cv,Morisita:1990cg}
$$
S_m={g_m \over 3}\begin{pmatrix}
1 & 0 & 0\\
0 & 1 & 0\\
0 & 0 & -2\\
\end{pmatrix} \,,
$$
where $g_m$ is the effective coupling constant. This matrix describes the mass breaking effect due to the $s$ and $u,d$ quarks mass difference related to the term
$$
{2 m_u+m_s \over 3} \overline q q + {m_u-m_s \over \sqrt{3}} \overline q \lambda_8 q \,,
$$
where the SU(2) isospin symmetry is assumed, so that: $m_u=m_d$. The $S_m$ matrix is proportional to the $8\mbox{-th}$ Gell-Mann matrix. The EM breaking effect is related to the fact that the photon-quarks coupling constant is proportional to the electric charge. This effect can be parametrized using the spurion matrix
$$
S_e={g_e \over 3}\begin{pmatrix}
2 & 0 & 0\\
0 & -1 & 0\\
0 & 0 & -1\\
\end{pmatrix} \,,
$$
where $g_e$ is the EM effective coupling constant.
\\
The most general SU(3)-invariant effective Lagrangian density, which accounts for these effects, is
\begin{eqnarray}
\label{eq.dens.lagr}
\mathcal L &=& g \, {\rm Tr}(\bb) + d \, {\rm Tr}\big(\{\b, \overline \b \} S_e \big)+f \, {\rm Tr}\big([\b, \overline \b ] S_e \big) \nonumber \\
&&+ d' \, {\rm Tr}\big(\{\b, \overline \b \} S_m \big)+f' \, {\rm Tr}\big([\b, \overline \b ] S_m \big) \,, 
\end{eqnarray}
where $g,d,f,d',f'$ are coupling constants. 
\\
By considering single $\bb$ final states, the complete Lagrangian density can be written as the sum of seven contributions
$$
\mathcal L = \mathcal L_{\Sigma^0 \Lambda} + \mathcal L_{p} + \mathcal L_{n} + \mathcal L_{\Sigma^+} + \mathcal L_{\Sigma^-} + \mathcal L_{\Xi^0} + \mathcal L_{\Xi^-} \,,
$$
with
\begin{eqnarray}
\mathcal L_{\Sigma^0 \Lambda} &=& \left(g+{1 \over 3}dg_e+{2 \over 3}d'g_m \right) \Sigma^0 \overline \Sigma^0 + \left({\sqrt{3} \over 3} d g_e \right) \Sigma^0 \overline \Lambda \nonumber \\
&&+ \left(g-{1 \over 3}dg_e-{2 \over 3}d'g_m \right) \Lambda \overline \Lambda + \left({\sqrt{3} \over 3} d g_e \right) \Lambda \overline \Sigma^0 \,, \nonumber \\
\mathcal L_{p} &=& \left( g+{1 \over 3}d g_e+fg_e-{1 \over 3}d'g_m+f'g_m \right) p \overline p \,, \nonumber \\
\mathcal L_{n} &=& \left( g-{2 \over 3}d g_e-{1 \over 3}d'g_m+f'g_m \right) n \overline n \,, \nonumber \\
\mathcal L_{\Sigma^+} &=& \left( g+{1 \over 3}d g_e+fg_e+{2 \over 3}d'g_m \right) \Sigma^+ \overline \Sigma^- \,, \nonumber \\
\mathcal L_{\Sigma^-} &=& \left( g+{1 \over 3}d g_e-fg_e+{2 \over 3}d'g_m \right) \Sigma^- \overline \Sigma^+ \,, \nonumber \\
\mathcal L_{\Xi^0} &=& \left( g-{2 \over 3}d g_e-{1 \over 3}d'g_m - f'g_m \right) \Xi^0 \overline \Xi^0 \,, \nonumber \\
\mathcal L_{\Xi^-} &=& \left( g+{1 \over 3}d g_e-fg_e-{1 \over 3}d'g_m - f'g_m \right) \Xi^- \overline \Xi^+ \,. \nonumber
\end{eqnarray}
Such a Lagrangian density allows us to parametrize the strong and the EM sub-amplitudes for the decays $J/\psi\to\bb$. In the next section, we will introduce an additional term to account also for the contribution due to the mixed strong-EM intermediate state.
\section{Amplitudes and sub-amplitudes}
The BR for the decay of the $J/\psi$ meson into a baryon-antibaryon pair \bb\ can be written as
$$
\br_{\bb} = \frac{\lmo\vec{p}\,\rmo}{8\pi  M_{J/\psi}^2 \Gamma_{J/\psi}}\left|\mathcal A^{ggg}_{\bb} + \mathcal A_{\bb}^{gg \gamma} + \mathcal A_{\bb}^\gamma\right|^2 \,,
$$
where $\vec{p}$ is the three-momentum of the baryon (antibaryon) in the $\bb$ center of mass frame ($J/\psi$ rest frame) and, $M_{J/\psi}$ and $\Gamma_{J/\psi}$ are the mass and the width of the $J/\psi$ meson. 
\\
As a consequence of the amplitude decomposition, the BR written as the sum of four contributions, i.e., 
\begin{equation}
\br_{\bb} \equiv \br^{ggg}_{\bb} + \br^{gg\gamma}_{\bb} + \br^{\gamma}_{\bb} + \br^{\rm int} \,,
\label{eq.strutturaBR}
\end{equation}
where the symbol $\br_{\bb}^{\sigma}$ stands for the BR due to the $\sigma$ intermediate state, with $\sigma=ggg,\,gg\gamma,\,\gamma$, while the last term accounts for the interference among the sub-amplitudes. 
\\
As already stated, this term depends only on the relative phase $\varphi$ between the sub-amplitudes $\big(\aggg_{\bb}+\agg_{\bb}\big)$ and $\ag_{\bb}$, because we assume that purely strong, $\aggg_{\bb}$, and the mixed strong-EM amplitude, $\agg_{\bb}$, have the same phase, i.e., they are relatively real and positive~\cite{Korner:1986vi}. Using the effective Lagrangian density of Eq.~\eqref{eq.dens.lagr}, the total decay amplitude can be parametrized in terms of the phase $\varphi$ and the coupling constants belonging to the set $\mathcal{C}=\{G_0,D_e,D_m,F_e,F_m\}$. Such five coupling constants are linear combinations of those appearing explicitly as coefficients of the traces in Eq.~\eqref{eq.dens.lagr}, i.e., $g,\,d,\,f,\,d',\,f'$.
\\
In particular, $G_0$ is related to the coupling constant $g$, $D_e,\,F_e$ to $d,\,g$, which refer to the EM breaking effects, and $D_m,\,F_m$ to $d',\,g'$, describing to the mass difference breaking effects.
\\
We assume that the strong and EM coupling constants of the subsets $\mathcal{C}_{\rm strong}=\{G_0,\,D_m,\,F_m\}$ and $\mathcal{C}_{\rm EM}~=~\{D_e,\,F_e\}$, respectively, are among each other relatively real and positive~\cite{Zhu:2015bha}. It follows that the only non-zero relative phase, that we call $\varphi$, is the one between strong and EM interactions. Moreover, since such a phase is mainly due to the dominant coupling constant $G_0$, we assume that $\varphi$ does not depend on the \bb\ final state.
\\
To account for the presence of the mixed strong-EM sub-amplitude $\agg_{\bb}$, we include the parameter $R$, that represents the ratio between the mixed strong-EM and the purely strong sub-amplitude
\begin{equation}
R = {\mathcal A^{gg \gamma}_{\bb} \over \mathcal A^{ggg}_{\bb}}\,.\label{eq:ratio}
\end{equation}
We assume that the dependence on the hadronization processes of the two-gluon-plus-one-photon and the three-gluon intermediate states cancels out in the ratio of the two sub-amplitudes, so that it has the same value for each \bb\ final state, likewise the coupling constants of the set $\mathcal C$.
Being proportional to the baryon charge~\cite{Claudson:1981fj,Chernyak:1984bm}, the sub-amplitude $\agg_{\bb}$, and hence the parameter $R$ are non-vanishing only for charged baryons.
\\
Asymptotically, i.e., at $q^2\gg\Lambda^2_{\rm QCD}$, when QCD is in perturbative regime (pQCD), the ratio $R$ scales as the ratio of the EM to the strong coupling constant~\cite{Korner:1986vi}
\begin{equation}
R_{\rm pQCD} (q^2)
= -{4 \over 5} {\alpha \over \alpha_S(q^2)} \,.
\label{eq:RpQCD}	
\end{equation}
Since, as already discussed, the realization of the perturbative regime at the $J/\psi$ mass is not well established, the value $R_{\rm pQCD}$ is not a good approximation for $R$.
\\
In light of that, the amplitude $\mathcal{A}_{\bb}$ for a given decay $J/\psi\to\bb$ can be written as combination of the coupling constants of the set $\mathcal{C}=\{G_0,D_e,D_m,F_e,F_m\}$, 
$$
\mathcal{A}_{\bb}=\sum_{\chi\in\mathcal{C}} c_{\bb}
^\chi\, \chi\,.
$$
where the coefficients $c_{\bb}
^\chi$ can assume the values of the set
$$
\{ 0,\sqrt{3},\pm 1,\pm 2,\pm e^{i\varphi},\pm 2e^{i\varphi}, Re^{i\varphi}, 2Re^{i\varphi} \}\,.
$$
The amplitudes for the decays $J/\psi\to\bb$ for the nine baryon-antibaryon pair of the spin-1/2 SU(3) octet are reported in Table~\ref{tab:A.par}.
\begin{table} [ht!]
\vspace{-2mm}
\centering
\caption{Amplitudes parameterization.}
\label{tab:A.par} 
\begin{tabular}{ll} 
\hline\noalign{\smallskip}
\bb & $\mathcal A_{\bb}=\mathcal A^{ggg}_{\bb} + \mathcal A^{gg\gamma}_{\bb} + \mathcal A^\gamma_{\bb}$ \\
\noalign{\smallskip}\hline\noalign{\smallskip}
$\Sigma^0 \overline \Sigma^0$ & $(G_0 + 2 D_m) e^{i \varphi} + D_e$ \\
$ \Lambda \overline \Lambda$ & $(G_0 - 2 D_m) e^{i \varphi} - D_e$ \\
$\Lambda \overline \Sigma^0 +$ c.c. & $\sqrt{3}\,D_e$ \\
$p \overline p$ & $(G_0 - D_m + F_m)(1+R) e^{i \varphi} + D_e + F_e$ \\
$n \overline n$ & $(G_0 - D_m + F_m) e^{i \varphi} - 2\,D_e$ \\
$\Sigma^+ \overline \Sigma^-$ & $(G_0 + 2 D_m)(1+R) e^{i \varphi} + D_e + F_e$ \\
$\Sigma^- \overline \Sigma^+$ & $(G_0 + 2 D_m)(1+R) e^{i \varphi} + D_e - F_e$ \\
$\Xi^0 \overline \Xi^0$ & $(G_0 - D_m - F_m) e^{i \varphi} - 2\,D_e$ \\
$\Xi^- \overline \Xi^+$ & $(G_0 - D_m - F_m)(1+R) e^{i \varphi} + D_e - F_e$ \\
\noalign{\smallskip}\hline
\end{tabular}
\end{table}\\
Sub-amplitudes can be easily identified. The purely EM $\ag_{\bb}$ is the combination of the coupling constants of the subset $\mathcal{C}_{\rm EM}$. The purely strong $\aggg_{\bb}$ is the combination of the coupling constants of the subset $\mathcal{C}_{\rm strong}$ multiplied by the phase $e^{i\varphi}$. Finally, the mixed strong-EM sub-amplitude $\agg_{\bb}$, which is present only for charged baryons, is given by $\agg_{\bb}=R\,\aggg_{\bb} $. For instance, in case of the decay $J/\psi\to p\overline{p}$, i.e., $\bb=p\overline{p}$, from the fourth row of Table~\ref{tab:A.par}, we have the sub-amplitudes
\begin{eqnarray}
\ag_{p\overline{p}} & = & D_e+F_e\,, \nonumber	\\
\aggg_{p\overline{p}} & = & \left( G_0-D_m+F_m\right) e^{i\varphi}\,,\nonumber\\
\agg_{p\overline{p}} & = & R\left( G_0-D_m+F_m\right) e^{i\varphi}\,.\nonumber
\end{eqnarray}
The neutron-antineutron, fifth row of Table~\ref{tab:A.par}, has the same purely strong sub-amplitude, different purely EM and vanishing mixed sub-amplitude, so that
$$
\aggg_{n\overline{n}}=\aggg_{p\overline{p}}
\,,\hspace{4mm}
\ag_{n\overline{n}}=-2D_e
\,,\hspace{4mm}
\agg_{n\overline{n}}=0\,.
$$
{\color{black} 
In general, the total amplitude for the $
\jp\to\bb$ decay is parametrized as
\begin{eqnarray}
\mathcal{A}_{\bb}
=\aggg_{\bb}(1+R)e^{i\varphi}+\ag_{\bb}
\equiv
\mathcal{S}_{\bb}\,e^{i\varphi}+\ag_{\bb}
\,,\label{eq:simple-form0}
\end{eqnarray}
where $\mathcal{S}_{\bb}=\aggg_{\bb}(1+R)$ and $\ag_{\bb}$ are real quantities to be determined by a minimization procedure, fitting the model predictions on the BRs to the corresponding experimental values. The total amplitude is defined up to an arbitrary, ineffective overall phase; by setting this phase to have $\mathcal{S}_{\bb}$ always positive, hence $\mathcal{S}_{\bb}=|\mathcal{S}_{\bb}|$, the sub-amplitude $\ag_{\bb}$ could be positive or negative, i.e., $\ag_{\bb}=|\ag_{\bb}|$ or $\ag_{\bb}=|\ag_{\bb}|e^{\pm i\pi}$. So that, the amplitude can be redefined up to an overall sign as
\be
\mathcal{A}_{\bb} \to \mathcal{A}_{\bb}=
|\mathcal{S}_{\bb}|e^{i \varphi_{\bb}}+|\ag_{\bb}|\,,
\label{eq:simple-form}
\en
where $\varphi_{\bb}=\varphi$ if $\ag_{\bb}>0$ and $\varphi_{\bb}=\varphi\pm\pi$ if $\ag_{\bb}<0$.
This form is useful in order to make comparisons with the moduli of sub-amplitudes and the relative phase that the BESIII Collaboration~\cite{Ablikim:2012eu} has obtained by fitting the data with a phenomenological parameterization of the amplitude.
The choice between $\varphi_{\bb}=\varphi+\pi$ and $\varphi_{\bb}=\varphi-\pi$, when $\ag_{\bb}$ is negative, is guided by the request that the total relative phase has to be in a given determination, for instance, $\varphi_{\bb}\in[0,2\pi]$. Actually, since the  experimental observable is the modulus squared of the amplitude $\mathcal{A}_{\bb}$, which depends only on the cosine of the relative phase, being
\be
\left| \mathcal{A}_{\bb}\right|^2=
|\mathcal{S}_{\bb}|^2+|\ag_{\bb}|^2
+2|\mathcal{S}_{\bb}||\ag_{\bb}|\cos\lt
\varphi_{\bb}\rt\,,
\nen
the ambiguity between the two values $\varphi^{\rm exp}_{\bb}$ and $2\pi-\varphi_{\bb}^{\rm exp}$, with $\varphi_{\bb}^{\rm exp}\in[0,\pi]$ and hence $2\pi-\varphi_{\bb}^{\rm exp}\in[\pi,2\pi]$, cannot be resolved. In other words, both values $\pm \varphi_{\bb}$, $|\mathcal{S}_{\bb}|$ and $|\ag_{\bb}|$ being equal, give the same modulus squared, because $|\mathcal{A}_{\bb}|^2=|\mathcal{A}^*_{\bb}|^2$,
where $\mathcal{A}^*_{\bb}$ is the complex conjugate of $\mathcal{A}_{\bb}$.
} 
\begin{table}[h]
\vspace{-2mm}
\centering
\caption{Branching ratios data from PDG~\cite{pdg} and BESIII experiment~\cite{Ablikim:2017tys}.}
\label{tab:BRJpsi.pdg} 
\begin{tabular}{lrr} 
\hline\noalign{\smallskip}
Decay process & Branching ratio & Error \\
\noalign{\smallskip}\hline\noalign{\smallskip}
$J/\psi \to \Sigma^0 \overline \Sigma^0$ & $(1.164 \pm 0.004) \times 10^{-3}$ & $0.34 \%$ \\
$J/\psi \to \Lambda \overline \Lambda$ & $(1.943 \pm 0.003) \times 10^{-3}$ & $0.15 \%$ \\
$J/\psi \to \Lambda \overline \Sigma^0 + {\rm c.c.} $ & $(2.83 \pm 0.23) \times 10^{-5}$ & $8.13 \%$ \\
$J/\psi \to p \overline p$ & $(2.121 \pm 0.029) \times 10^{-3}$ & $1.37 \%$ \\
$J/\psi \to n \overline n$ & $(2.09 \pm 0.16) \times 10^{-3}$ & $7.66 \%$ \\
$J/\psi \to \Sigma^+ \overline \Sigma^-$ & $(1.50 \pm 0.24) \times 10^{-3}$ & $16.00 \%$ \\
$J/\psi \to \Xi^0 \overline \Xi^0$ & $(1.17 \pm 0.04) \times 10^{-3}$ & $3.42 \%$ \\
$J/\psi \to \Xi^- \overline \Xi^+$ & $(9.7 \pm 0.8) \times 10^{-4}$ & $8.25 \%$ \\
\noalign{\smallskip}\hline
\end{tabular}
\end{table}\\%
\section{Electromagnetic couplings and $\chi^2$ definition}
Data are available for eight of the nine decays, in particular, the decay $J/\psi\to\Sigma^+\overline{\Sigma}^-$ is the only one that has not yet been observed. All available data are reported in Table~\ref{tab:BRJpsi.pdg}. \\
From the BR of the decay $J/\psi \to (\Lambda \overline \Sigma^0 + {\rm c.c.})$, which is purely electromagnetic, we can extract the modulus of $D_e$ as
$$
|D_e|=\sqrt{16 \pi M_{J/\psi} \Gamma_{J/\psi}\br(J/\psi \to \big(\Lambda \overline \Sigma^0 + {\rm c.c.}\big) \over 3 \beta_{\Lambda \overline \Sigma^0}} \,,
$$
where the outgoing baryon velocity in the $J/\psi$ center of mass is defined as
$$
\beta_{\Lambda \overline \Sigma^0} \equiv \sqrt{1-{2(M_{\Sigma^0}^2+M_{\Lambda}^2) \over M_{J/\psi}^2} + {(M_{\Sigma^0}^2-M_{\Lambda}^2)^2 \over M_{J/\psi}^4}} \,.
$$
Using the experimental value of BR$_{\Lambda \overline \Sigma^0}$, given in the third row of Table~\ref{tab:BRJpsi.pdg}, we obtain
$$
|D_e|=(4.52 \pm 0.20) \times 10^{-4} \ \rm GeV \,.
$$
Analogously, the EM BR of the decay of the $J/\psi$ meson into proton-antiproton is given by
$$
\br^\gamma_{p \overline p} = {\beta_{p \overline p} \over 16 \pi M_{J/\psi}\Gamma_{J/\psi}} |D_e+F_e|^2\,,
$$
where the proton velocity is
$$
\beta_{p \overline p}=\sqrt{1-\frac{4M_p^2}{M_{J/\psi}^2}}\,.
$$
The EM BR is related to the $e^+e^-\to p\overline{p}$ non-resonant cross section at the $J/\psi$ mass by the formula~\cite{Ferroli:2016jri}
\begin{equation}
\label{eq.BRgamma}
\br^\gamma_{\bb} = \br_{\mu\mu} \, {\sigma_{\ee \to \bb}(M_{\jp}^2) \over \sigma^0_{\ee \to \mu^+ \mu^-}(M_{\jp}^2)} \,,
\end{equation}
where $\br_{\mu\mu}$ is the BR of the decay $J/\psi \to \mu^+\mu^-$, and $\sigma^0_{\ee \to \mu^+ \mu^-}(q^2)$ represents the bare $\ee \to \mu^+ \mu^-$ cross section, i.e., the cross section corrected for the vacuum-polarization 
$$
\sigma^0_{\ee \to \mu^+ \mu^-} (q^2)= {4 \pi \alpha^2 \over 3 q^2} \,.
$$
The modulus of the sum of the two parameters $D_e$ and $F_e$ has, therefore, the expression 
$$
|D_e + F_e| = \sqrt{12 \, \br_{\mu\mu} M_{J/\psi}^3 \Gamma_{J/\psi} \,\sigma_{\ee \to p \overline p}(M_{\jp}^2) \over \alpha^2 \beta_{p \overline p}} \,.
$$
By using the value
\begin{eqnarray}
\sigma_{\ee\to p \overline p} (M_{\jp}^2)&=& {6912 \,\pi \alpha^2 (M_{J/\psi}^2\!\!+\!2M_p^2) \over M_{J/\psi}^{12}\,{\rm GeV^{-8}}} 
\nonumber\\
&&
\times\!\!\left[ \ln^2\left(\!{M_{J/\psi}^2 \over 0.52^2\,\rm GeV^2}\!\right)\!+\!\pi^2 \right]^{\!\!-2} ,
\end{eqnarray}
for the $\ee \to p \overline p$ cross section at $q^2=M_{J/\psi}^2$, that has been obtained by the most recent measurement of the BESIII Collaboration~\cite{Ablikim:2019njl}, and~\cite{pdg}
$$
\br_{\mu\mu} = (5.961 \pm 0.033) \times 10^{-2} \,,
$$
for the $\mu^+\mu^-$ BR, we obtain
\begin{equation}
\label{eq.brEMpp}
\br^{\gamma,\rm exp}_{p\overline{p}} = (8.46 \pm 0.79) \times 10^{-5}
\end{equation}
and
$$
|D_e + F_e| = (1.240 \pm 0.061) \times 10^{-3} \ \rm GeV \,.
$$
The error includes both statistical and systematic contributions due to the cross section fit to the BESIII data.
\\
We define the $\chi^2$ as follows
\begin{eqnarray}
\label{eq.chi.2}
\chi^2\left(\mathcal{C};R,\varphi\right) &=& \sum_{\bb} \left({\br_{\bb}^{\rm th}-\br_{\bb}^{\rm exp} \over \delta \br_{\bb}^{\rm exp}}\right)^2
\nonumber\\ &&+
\left({\br_{p\overline{p}}^{\gamma,\rm th}-\br_{p\overline{p}}^{\gamma,\rm exp} \over \delta \br_{p\overline{p}}^{\gamma,\rm exp}}\right)^2
\end{eqnarray}
where the sum runs over the eight baryon-antibaryon pairs, \bb, for which experimental data are available, reported in Table~\ref{tab:BRJpsi.pdg}, the second term imposes the constraint of the EM BR reported in Eq.~\eqref{eq.brEMpp}. 
\\
The numerical minimization is performed with respect to the five coupling constants of the set  $\mathcal C=\{G_0,D_e,D_m,F_e,F_m\}$, the ratio $R$ defined in Eq.~\eqref{eq:ratio}, and the relative phase $\varphi$.
\section{Results}
The best values of the parameters resulting from the numerical minimization of the $\chi^2$ defined in Eq.~\eqref{eq.chi.2} are shown in Table~\ref{tab:results.par}. The errors have been obtained by means of a Monte-Carlo Gaussian simulation.
\begin{table} [h]
\vspace{-2mm}
\centering
\caption{Values of the parameters from the $\chi^2$ minimization.}
\label{tab:results.par} 
\begin{tabular}{lr} 
\hline\noalign{\smallskip}
$G_0$ & $(5.73511 \pm 0.0059) \times 10^{-3} \ \rm GeV$ \\
$D_e$ & $(4.52 \pm 0.19) \times 10^{-4} \ \rm GeV$ \\
$D_m$ & $(-3.74 \pm 0.34) \times 10^{-4} \ \rm GeV$ \\
$F_e$ & $(7.91 \pm 0.62) \times 10^{-4} \ \rm GeV$  \\
$F_m$ & $(2.42 \pm 0.12) \times 10^{-4} \ \rm GeV$ \\
$\varphi$ & $1.27 \pm 0.14 = (73 \pm 8)^\circ $  \\
$R$ & $(-9.7 \pm 2.1) \times 10^{-2}$ \\
\noalign{\smallskip}\hline
\end{tabular}
\end{table}\\
The BRs are reported in Table~\ref{tab:BRcalc.J.psi}, where they are compared with the corresponding experimental values.  It is interesting to notice that the obtained value for the BR of the unobserved $J/\psi \to \Sigma^- \overline \Sigma^+$ decay represents a prediction of the model.
\begin{figure}[h]
	\begin{center}
		\includegraphics[width=\columnwidth]{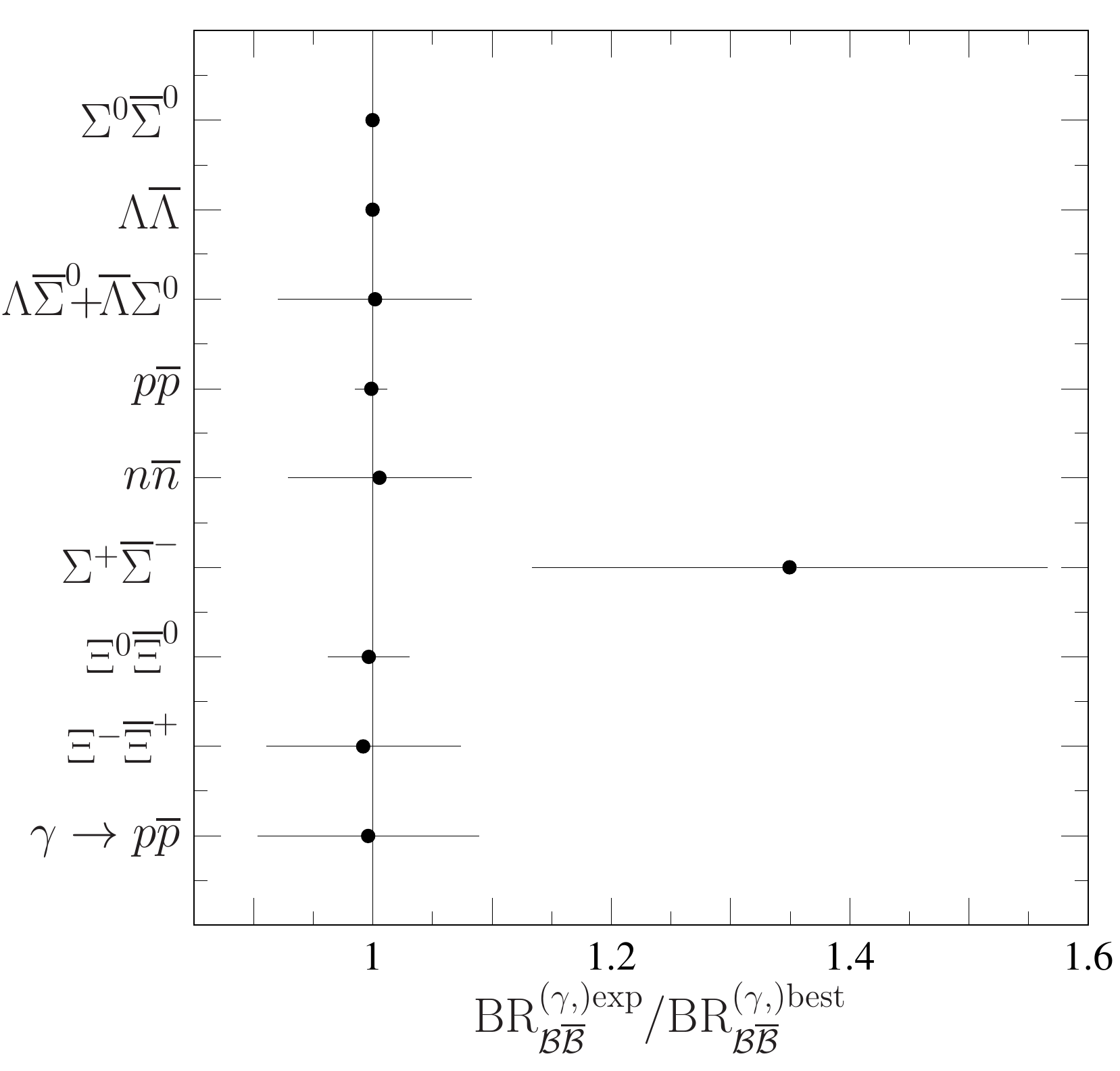}
		\caption{Ratios between the experimental input values of BRs and their best values obtained by minimizing the $\chi^2$ of Eq.~\eqref{eq.chi.2}. The lower point, at the ordinate labelled width $\gamma\to p\overline{p}$, is the contribution due to EM BR of the proton, see Eq.~\eqref{eq.brEMpp}.}\label{fig:rchi2}
	\end{center}
\end{figure}
\\
Figure~\ref{fig:rchi2} shows the ratios between the input and best values of the BRs that represent the nine free parameters of the $\chi^2$ given in Eq.~\eqref{eq.chi.2}. The minimum normalized $\chi^2$ is
\be
\frac{\chi^2\left(\mathcal{C}^{\rm best};R^{\rm best},\varphi^{\rm best}\right)}{N_{
\rm dof}} &=& 1.33\,,
\label{eq:chi2}\en
where the number of degrees of freedom is $N_{\rm dof}=N_{\rm const}-N_{\rm param}=2$, having nine constraints, $N_{\rm const}=9$ and seven free parameters, $N_{\rm param}=7$. 
\\
%
%
{\color{black}  
It is interesting to notice that the significance of the mixed strong-EM contribution in the description of the \jp\ decay mechanism can be verified by comparing the normalized $\chi^2$'s obtained in the case where $R$ is considered as a free parameter, Eq.~\eqref{eq:chi2}, to that in which it is fixed at $R=0$, i.e.,
\be
\frac{\chi^2\left(\mathcal{C'}^{\rm best};R=0,\varphi'^{\rm best}\right)}{N_{
\rm dof}}=\frac{16.44}{3} &=& 5.48\,,
\label{eq:chi2-2}\en
where $\mathcal{C'}^{\rm best}$ and $\varphi'^{\rm best}$ are the set of best values of the coupling constants and the best relative phase obtained in this case. Despite the quite low number of degrees of freedom\footnote{\color{black}%
%
%
The most suitable criterion to compare these two hypotheses, namely: $R$ free and $R=0$, is the one provided by the $p$-value, $p(\chi^2;N_{\rm dof})$. The two values are
$$
p(2.65;2)=0.266\,,\hh
p(16.44;3)=9.21\times 10^{-4}\,,
$$ 
and represent the probabilities to obtain by chance $\chi^2=2.65$ and $\chi^2=16.44$, with two and three degrees of freedom respectively, if the model is correct.
}%
%
and also the smallness of the best value obtained for $R$, see Table~\ref{tab:results.par}, this large $\chi^2$ value, Eq.~\eqref{eq:chi2-2}, represents a clear indication in favor of the necessity of the mixed EM-strong contribution.   
} 
\\
 The knowledge of the seven coupling constants of Table~\ref{tab:results.par} brings important information on the structure of the amplitudes for the baryonic decays of the $J/\psi$ meson. Indeed, using these values under the assumptions concerning their relative phases, individual sub-amplitudes can be calculated. Table~\ref{tab:BRsepcalc.J.psi} reports the purely strong, purely EM and mixed strong-EM contribution to the total BR for the nine final states. The value of the ratios between the moduli of the sub-amplitudes $|\mathcal A^\gamma_{\bb}/ \mathcal A^{ggg}_{\bb}|$ and $|\mathcal A^{gg\gamma}_{\bb}/ \mathcal A^{ggg}_{\bb}|$ are shown in Table~\ref{tab:BRsepcalc.J.psi.ratio}. 
 \\
 {\color{black} 
 The moduli of the sub-amplitudes $\mathcal{S}_{\bb}$ and $\ag_{\bb}$ together with the phase $\varphi_{\bb}$, defined in Eqs.~\eqref{eq:simple-form0} and~\eqref{eq:simple-form}, are reported in Table~\ref{tab:simple-form}. The five final states: $\Lambda\ov{\Lambda}$, $n\ov{n}$, $\Sigma^-\ov{\Sigma}^+$, $\Xi^0\ov{\Xi}^0$, $\Xi^-\ov{\Xi}^+$, have negative $\ag_{\bb}$ sub-amplitudes and then $\varphi_{\bb}=\pi-\varphi$. This is a phenomenological finding. It is due to the values that have been obtained for the coupling constants $D_e$ and $F_e$ with the fitting procedure, see Table~\ref{tab:results.par}, and to the SU(3) symmetry of the model, that determines the signs of the coupling constants in the definition of the sub-amplitudes, see Table~\ref{tab:A.par}. As an example, let us consider the $p\ov{p}$ and $n\ov{n}$ final states. Using the standard parameterization of Eq.~\eqref{eq:simple-form}, the total relative phase between the two sub-amplitudes $\mathcal{S}_{\bb}e^{i\varphi}$ and $\ag_{\bb}$ differ by $180^\circ$, i.e.,
 \be
 \arg\lt\frac{\mathcal{S}_{p\ov{p}}e^{i\varphi}}{\ag_{p\ov{p}}}\rt=\varphi\,,\hh \arg\lt\frac{\mathcal{S}_{n\ov{n}}e^{i\varphi}}{\ag_{n\ov{n}}}\rt=\varphi\pm\pi\,.
 \nen
 The last result is a consequence of the negative value of $\ag_{n\ov{n}}=-2D_e$, with $D_e>0$, see Table~\ref{tab:results.par}.
 } 
\begin{table}[h]
\vspace{-2mm}
\centering
\caption{Branching ratios from PDG~\cite{pdg} (second column), from parameters of Table~\ref{tab:results.par} (third column) end their difference in units of the total error (fourth column).}
\label{tab:BRcalc.J.psi} 
\begin{tabular}{lrrr} 
\hline\noalign{\smallskip}
\bb & BR$^{\rm PDG}_{\bb}\times 10^3$ & BR$_{\bb}\times 10^3$ & $\Delta {\br} / \sum \sigma_\br$ \\
\noalign{\smallskip}\hline\noalign{\smallskip}%
$\Sigma^0 \overline \Sigma^0$ & $1.164 \pm 0.004 $ & $1.160 \pm 0.041 $ & $\sim 0.09$ \\
$\Lambda \overline \Lambda$ & $1.943 \pm 0.003 $ & $1.940 \pm 0.055 $ & $\sim 0.05$ \\
$\Lambda \overline \Sigma^0 +$ c.c. & $0.0283 \pm 0.0023$ & $0.0280 \pm 0.0024$ & $\sim 0.06$ \\
$p \overline p$ & $2.121 \pm 0.029 $ & $2.10 \pm 0.16 $ & $\sim 0.1$ \\
$n \overline n$ & $2.09 \pm 0.16 $ & $2.10 \pm 0.12 $ & $\sim 0.04$ \\
$\Sigma^+ \overline \Sigma^-$ & $1.50 \pm 0.24 $ & $1.110 \pm 0.086 $ & $\sim 1$ \\
$\Sigma^- \overline \Sigma^+$ & $/$ & $0.857 \pm 0.051 $ & $/$ \\
$\Xi^0 \overline \Xi^0$ & $1.17 \pm 0.04 $ & $1.180 \pm 0.072 $ & $\sim 0.09$ \\
$\Xi^- \overline \Xi^+$ & $0.97 \pm 0.08 $ & $0.979 \pm 0.065 $ & $\sim 0.06$ \\
\noalign{\smallskip}\hline
\end{tabular}
\end{table}%
\begin{table}
\vspace{-2mm}
\centering
\caption{Purely strong (second column), purely EM (third column) and mixed (fourth column) BRs.}
\label{tab:BRsepcalc.J.psi} 
\begin{tabular}{lrrr} 
\hline\noalign{\smallskip}
\bb & $\br^{ggg}_{\bb} \times 10^{3}$ & $\br^{\gamma}_{\bb} \times 10^{5}$ & $\br^{gg\gamma}_{\bb} \times 10^{5}$ \\
\noalign{\smallskip}\hline\noalign{\smallskip}%
$\Sigma^0 \overline \Sigma^0$            & $1.100 \pm 0.030$ & $0.902 \pm 0.076$       & $0$             \\
$\Lambda \overline \Lambda$              & $2.020 \pm 0.042$ & $0.981 \pm 0.083$       & $0$             \\
$\Lambda \overline \Sigma^0+$ c.c.  & $0$               & $2.83 \pm 0.24$         & $0$             \\
$p \overline p$                          & $2.220 \pm 0.085$ & $8.52 \pm 0.89$         & $2.19 \pm 0.93$ \\
$n \overline n$                          & $2.220 \pm 0.085$ & $4.50 \pm 0.38$         & $0$             \\
$\Sigma^+ \overline \Sigma^-$            & $1.100 \pm 0.030$ & $6.86 \pm 0.72$         & $1.08 \pm 0.46$ \\
$\Sigma^- \overline \Sigma^+$            & $1.090 \pm 0.030$ & $0.52 \pm 0.20$         & $1.07 \pm 0.46$ \\
$\Xi^0 \overline \Xi^0$                  & $1.260 \pm 0.053$ & $2.99 \pm 0.25$         & $0$             \\
$\Xi^- \overline \Xi^+$                  & $1.240 \pm 0.052$ & $0.43 \pm 0.16$         & $1.22 \pm 0.52$ \\
\noalign{\smallskip}\hline
\end{tabular}
\end{table}
\begin{table}
\vspace{-2mm}
\centering
\caption{Approximate values of moduli of the ratios between sub-amplitudes $\mathcal A^\gamma_{\bb}$ and $\mathcal A^{ggg}_{\bb}$ (second column), and between $\mathcal A^{gg\gamma}_{\bb}$ and $\mathcal A^{ggg}_{\bb}$ (third column).}
\label{tab:BRsepcalc.J.psi.ratio} 
\begin{tabular}{lrr} 
\hline\noalign{\smallskip}
\bb $ \ $ & $|\mathcal A^\gamma_{\bb}/ \mathcal A^{ggg}_{\bb}| $ & $|\mathcal A^{gg\gamma}_{\bb}/ \mathcal A^{ggg}_{\bb}| $ \\
\noalign{\smallskip}\hline\noalign{\smallskip}%
$\Sigma^0 \overline \Sigma^0$ & $\sim 0.09$ & $0$ \\
$\Lambda \overline \Lambda$ & $\sim 0.07$& $0$ \\
$p \overline p$ & $\sim 0.20$ & $\sim 0.1$ \\
$n \overline n$ & $\sim 0.14$ & $0$ \\
$\Sigma^+ \overline \Sigma^-$ & $\sim 0.25$ & $\sim 0.1$ \\
$\Sigma^- \overline \Sigma^+$ & $\sim 0.07$ & $\sim 0.1$ \\
$\Xi^0 \overline \Xi^0$ & $\sim 0.15$ & $0$ \\
$\Xi^- \overline \Xi^+$ & $\sim 0.06$ & $\sim 0.1$ \\
\noalign{\smallskip}\hline
\end{tabular}
\end{table}\\%
\begin{table} 
\vspace{-2mm}
\centering
\caption{\color{black}
Moduli of sub-amplitudes $\mathcal{S}_{\bb}$, $\ag_{\bb}$ and phase $\varphi_{\bb}$, defined in Eq.~\eqref{eq:simple-form}.\label{tab:simple-form} 
}
{\color{black} 
\begin{tabular}{lcrcrcr} 
\hline\noalign{\smallskip}
\bb $ \ $ && $|\mathcal{S}_{\bb}|\times 10^3 $ && $|\ag_{\bb}|\times 10^4 $ && $\varphi_{\bb} $\\
\noalign{\smallskip}\hline\noalign{\smallskip}%
$\Sigma^0 \overline \Sigma^0$ && $4.987\pm 0.065$ && $4.52\pm0.19$ && $\varphi$\\
$\Lambda \overline \Lambda$ && $6.483\pm 0.065$ && $4.52\pm0.19$ && $\pi-\varphi$\\
$\Lambda\overline{\Sigma}^0+$ c.c. && 0 && $7.83\pm 0.33$ && $\varphi$\\
$p \overline p$ && $5.74\pm 0.14$ && $12.43\pm 0.65$ && $\varphi$\\
$n \overline n$ && $6.351\pm 0.037$ && $9.04\pm 0.38$ && $\pi-\varphi$\\
$\Sigma^+ \overline \Sigma^-$ && $4.50\pm 0.12$ && $12.43\pm 0.65$ && $\varphi$\\
$\Sigma^- \overline \Sigma^+$ && $4.50\pm 0.12$ && $3.39\pm 0.65$ && $\pi-\varphi$\\
$\Xi^0 \overline \Xi^0$ && $5.867\pm 0.037$ && $9.04\pm 0.38$ && $\pi-\varphi$\\
$\Xi^- \overline \Xi^+$ && $5.30\pm 0.13$ && $3.39\pm 0.65$ && $\pi-\varphi$\\
\noalign{\smallskip}\hline
\end{tabular}
}   
\end{table}\\%
Finally, we use the obtained values for the EM BRs (second column of Table~\ref{tab:BRsepcalc.J.psi}) to calculate non-resonant $\ee \to \bb$ Born cross sections at $q^2=M^2_{J/\psi}$, the results are reported in Table~\ref{tab:sigma.gamma}.
\\
The majority of these results does represents a prediction because currently there are no data on the corresponding processes. The only exception concerns BR$_{p \overline p}^\gamma$, whose experimental value, given in Eq.~\eqref{eq.brEMpp}, extracted from the non-resonant $\ee \to p \overline p$ cross section data~\cite{Ablikim:2019njl}, has been used as a constraint in the numerical $\chi^2$ minimization.
\begin{table} 
\vspace{-2mm}
\centering
\caption{Non-resonant $\ee \to \bb$ Born cross sections at $q^2=M_{J/\psi}^2$.}
\label{tab:sigma.gamma} 
\begin{tabular}{lr} 
\hline\noalign{\smallskip}
$\ee\to\bb$ & Cross section at the $q^2=M_{J/\psi}^2$ \\
\noalign{\smallskip}\hline\noalign{\smallskip}
$\ee \to \Sigma^0 \overline \Sigma^0$ & $(1.37 \pm 0.12) \ \rm pb$ \\
$\ee \to \Lambda \overline \Lambda$ & $(1.49 \pm 0.13) \ \rm pb$ \\
$\ee \to (\Lambda \overline \Sigma^0 + {\rm c.c.}) \ $ & $(4.30 \pm 0.36) \ \rm pb$ \\
$\ee \to p \overline p$ & $(12.9 \pm 1.4) \ \rm pb$ \\
$\ee \to n \overline n$ & $(6.84 \pm 0.58) \ \rm pb$ \\
$\ee \to \Sigma^+ \overline \Sigma^-$ & $(10.4 \pm 1.1) \ \rm pb$ \\
$\ee \to \Sigma^- \overline \Sigma^+$ & $(0.79 \pm 0.30) \ \rm pb$ \\
$\ee \to \Xi^0 \overline \Xi^0$ & $(4.54 \pm 0.38) \ \rm pb$ \\
$\ee \to \Xi^- \overline \Xi^+$ & $(0.65 \pm 0.24) \ \rm pb$ \\
\noalign{\smallskip}\hline
\end{tabular}
\end{table}\\%
\section{Conclusions}
We have developed a model based on an effective Lagrangian density to describe the decay amplitude of the $\jp$ meson into baryon-antibaryon pairs of the spin-1/2 SU(3) octet. It depends on seven free parameters: the five coupling constants of the set $\mathcal C=\{G_0,D_e,D_m,F_e,F_m\}$, the ratio $R$, that takes into account the mixed strong-EM contribution, and the phase $\varphi$ between the strong, $\aggg_{\bb}$, and the EM sub-amplitude, $\ag_{\bb}$ (the mixed sub-amplitude $\agg_{\bb}$ has the same phase of $\aggg_{\bb}$).
\\
The parameters have been determined by means of a fitting procedure, having as inputs nine experimental data: the eight BRs reported in Table~\ref{tab:BRJpsi.pdg}, and EM BR of the decay $J/\psi \to p \overline p$, extracted from the non-resonant $\ee \to p \overline p$ cross section at the $J/\psi$ mass, whose value is given in Eq.~\eqref{eq.brEMpp}.
\\
The best values of the parameters are shown in Table~\ref{tab:results.par}, while the corresponding BRs, together with the experimental inputs are listed in Table~\ref{tab:BRcalc.J.psi}. Notice that the result for BR$_{\Sigma^- \overline \Sigma^+}$ is actually a prediction since there are no data on the $\jp \to \Sigma^- \overline \Sigma^+$ decay rate.
\\
The model allows to determine the three single contributions (purely strong, purely EM and mixed strong-EM) to the total BR for all the \bb\ final states, their values are reported in Table~\ref{tab:BRsepcalc.J.psi}. Table ~\ref{tab:BRsepcalc.J.psi.ratio} shows, instead, the relative strengths of the purely EM and the mixed sub-amplitudes with respect to the purely strong one. Using the purely EM sub-amplitudes in the expression of Eq.~\eqref{eq.BRgamma}, we obtained the predictions for eight non-resonant $\ee \to\bb$ cross sections at the $\jp$ mass, such predictions, together with the proton cross section used as input, are reported in Table~\ref{tab:sigma.gamma}. 
\\
 The best value of the relative phase is
$$
\varphi = (73\pm 8)^\circ \,,
$$
it agrees with the result given in Refs.~\cite{Zhu:2015bha,Baldini:1998en} and with that given in Ref.~\cite{Ablikim:2012eu}, i.e., $(88.7 \pm 8.1)^\circ$, obtained by studying the decays of the $J/\psi$ meson into nucleon-antinucleon. 
{\color{black} 
More in detail, by considering the relative sign between the sub-amplitudes $|\mathcal{S}_{\bb}|$ and $|\ag_{\bb}|$, defined in Eq.~\eqref{eq:simple-form}, we can distinguish between two values of the relative phase $\varphi_{\bb}$, see Table~\ref{tab:simple-form},
\be
\varphi_{\Sigma^0\overline{\Sigma}^0,\Lambda\overline{\Sigma}^0,p\overline{p},\Sigma^+\overline{\Sigma}^-}&=&(73\pm 8)^\circ\,,
\no\\
\varphi_{\Lambda\overline{\Lambda},n\overline{n},\Sigma^-\overline{\Sigma}^+,\Xi^0\overline{\Xi}^0,\Xi^-\overline{\Xi}^+}&=&(107\pm 8)^\circ\,.
\nen}
As already discussed, the fact that the relative phase $\varphi$ is closer to $90^\circ$ rather than to $0^\circ$ or $180^\circ$ disagrees with pQCD. Indeed, in the perturbative regime, all QCD amplitudes should be real.  It follows that the obtained relative phase could be interpreted as the indication of a non-complete realization of pQCD at this energy, at least for these \jp\ decays. 
\\
Another result that points in the same direction is the value obtained for the ratio $R$, i.e.,
 $$
	R = -0.097\pm0.021\,.
 $$
In fact, such value, besides confirming that the mixed strong-EM sub-amplitude is negligible with respect to purely strong one, restates that at the $J/\psi$ mass the perturbative regime of QCD is still not reached. Indeed, it is not compatible with that predicted by pQCD itself, which is, see Eq.~\eqref{eq:RpQCD},
$$
 R_{\rm pQCD}(M_{J/\psi}^2) = -{4 \over 5} {\alpha \over \alpha_S(M_{J/\psi}^2)} \sim -0.030\,,
$$
where we have used $\alpha_S(M_{J/\psi}^2)\sim 0.2$~\cite{Claudson:1981fj}. %
{\color{black} 
A similar conclusion is also suggested Ref.~\cite{Chanowitz:1975ee}. %
}
\\
The possibility of disentangling single contributions allows, for the first time, to determine the mixed strong-EM sub-amplitude for each charged \bb\ final state. The third column of Table~\ref{tab:BRsepcalc.J.psi.ratio} reports the strength of such sub-amplitude relative to the dominant three-gluon one. In all cases, because by assumption it does not depend on the \bb\ final state, it represent about 10\% of the dominant contribution and becomes $\sim 1\%$ for the BR, see the fourth column of Table~\ref{tab:BRsepcalc.J.psi}. More intriguing is the comparison between the mixed and the purely EM contributions, third and fourth columns of Table~\ref{tab:BRsepcalc.J.psi} for the BRs and, second and third columns of Table~\ref{tab:BRsepcalc.J.psi.ratio} for the sub-amplitudes. They are always of the same order, but while for the proton and $\Sigma^+$ the modulus of the purely EM sub-amplitude is about twice the modulus of the mixed sub-amplitude, in the cases of $\Sigma^-$ and $\Xi^-$ the hierarchy is inverted. Such different behavior could be due to the different quark structure of the two pairs of baryons.
\\
Even though the model is based on an effective QCD Lagrangian and hence it is mainly devoted to describing the strong dynamics, it can also be exploited to predict non-resonant $\ee\to\bb$ cross sections at the \jp\ mass. By using the formula of Eq.~\eqref{eq.BRgamma}, the cross section $\sigma_{\ee\to\bb}(M_{\jp}^2)$ can be computed in terms of the purely EM BR, i.e., BR$^\gamma_{\bb}$. Apart from that of the proton, used as input, all the other cross section values represent predictions of the model. We would like to pay close attention to the neutron cross section, that probably will be measured soon. Our prediction, see fifth row of Table~\ref{tab:sigma.gamma},
$$
\sigma_{\ee\to n\overline{n}}(M_{\jp}^2)=(6.84\pm 0.58)\,{\rm pb}
$$
is in agreement with the natural expectation
\begin{eqnarray}
\sigma^{\rm expected}_{\ee\to n\overline{n}}(M_{\jp}^2)
&=&\left(\frac{\mu_n}{\mu_p}\right)^2\sigma_{\ee\to p\overline{p}}(M_{\jp}^2)
\nonumber\\
&=& (6.1\pm 0.7)\,{\rm pb}\,,\nonumber
\end{eqnarray}
which is obtained by scaling the proton cross section, reported in the fourth row of Table~\ref{tab:sigma.gamma}, by the squared  of the ratio between neutron, $\mu_n$, and proton, $\mu_p$, magnetic moment.
\section*{Acknowledgement}
We would like to warmly acknowledge the Italian group of the {BESIII} Collaboration for the useful and fruitful discussions on experimental and phenomenological aspects concerning \jp\ decays.

\end{document}